\documentclass[letterpaper,10pt,conference]{ieeeconf}  % Comment this line out if you need a4paper
\usepackage{amssymb}
\usepackage{color}
\usepackage{graphicx}
\usepackage{mathtools}
\usepackage{multirow}
\usepackage{lineno,hyperref}
\usepackage{url}
\usepackage{cleveref}
\usepackage{bm}
\usepackage{tikz}
\usepackage{algorithm}
\usepackage{algorithmicx}
\usepackage{algpseudocode}
\usepackage{caption}
\usepackage{pgfplots}

\modulolinenumbers[5]

\newtheorem{remark}{Remark}[section]

\usepackage{multirow}
\usepackage{color}
\usepackage{paralist}
\usepackage[space]{cite}

\newcommand{\ie}{\emph{i.e.}}
\newcommand{\eg}{\emph{e.g.}}
\newcommand{\R}{\mathbb{R}}
\newcommand{\E}{\mathbb{E}}
\renewcommand{\P}{\mathbb{P}}

\newcommand{\lrp}[1]{\left(#1\right)}
\newcommand{\lrs}[1]{\left[#1\right]}

\renewcommand{\t}{\theta}

\newcommand{\h}{\hat}

\newcommand{\ra}{\rightarrow}

\newcommand{\nn}{\nonumber}

\newcommand{\T}{\Theta}
\renewcommand{\d}{\delta}

\renewcommand{\i}[2]{\int_{#1}^{#2}}

\renewcommand{\S}[3]{\sum_{\substack{#1 \\ #2}}^{#3}}

\renewcommand{\l}{\label}

\pgfplotsset{compat=1.5}

\providecommand{\keywords}[1]{\textbf{\textit{Index terms---}}}

\IEEEoverridecommandlockouts                              % This command is only needed if 
\title{\LARGE \bf A Statistical Decision-Theoretical Perspective on the Two-Stage Approach to Parameter Estimation}

\author{Braghadeesh Lakshminarayanan and Cristian R. Rojas % <-this % stops a space
\thanks{This work has been supported by the Swedish Research Council under contract number 2016-06079 (NewLEADS) and by the Digital Futures project EXTREMUM. The authors are with the Division of Decision and Control Systems, KTH Royal Institute of Technology, 100 44 Stockholm, Sweden (e-mails: blak@kth.se; crro@kth.se).}%
}
\begin{document}

\maketitle

%%%%%%%%%%%%%%%%%%%%%%%%%%%%%%%%%%%%%%%%%%%%%%%%%%%%%%%%%%%%%%%%%%%%%%%%%%%%%%%%
\begin{abstract}
One of the most important problems in system identification and statistics is how to estimate the unknown parameters of a given model. Optimization methods and specialized procedures, such as Empirical Minimization (EM) can be used in case the likelihood function can be computed. For situations where one can only simulate from a parametric model, but the likelihood is difficult or impossible to evaluate, a technique known as the Two-Stage (TS) Approach can be applied to obtain reliable parametric estimates. Unfortunately, there is currently a lack of theoretical justification for TS. In this paper, we propose a statistical decision-theoretical derivation of TS, which leads to \emph{Bayesian} and \emph{Minimax} estimators. We also show how to apply TS on models for independent and identically distributed samples, by computing quantiles of the data as a first step, and using a linear function as the second stage. The proposed method is illustrated via numerical simulations.
\end{abstract}

\section{Introduction}

Engineering disciplines rely on mathematical models in order to understand, modify and control physical systems.
To derive such models, practitioners utilize techniques from several research fields, such as statistics~\cite{casella2021statistical}, chemometrics~\cite{deming1988chemometrics}, econometrics~\cite{Gourieroux-89}, machine learning~\cite{Shalev-Shwartz-Ben-David-12} and system identification~\cite{Ljung:99,Soderstrom-Stoica-89}, which build models based on data gathered from a physical system. 

Many models built from data are \emph{parametric statistical models}, \ie, they define (explicitly or implicitly) a probability distribution that describes the observed data in terms of a set of unknown parameters, which remain to be estimated. Many techniques have been devised for estimating such parameters, including Maximum Likelihood~\cite{casella2021statistical,LehmCase98}, Prediction Error Methods~\cite{Ljung:99}, Instrumental Variables~\cite{Soderstrom-Stoica-83}, Methods of Moments~\cite{Gourieroux-89}, and so on. While some of these techniques achieve statistical efficiency~\cite{casella2021statistical}, their implementations have several drawbacks: they may require having an explicit expression for the likelihood function, solving non-convex optimization problems, or that the distribution of the samples has a very specific structure. To address some of these issues, novel techniques have been introduced in econometrics and identification, such as Indirect Inference and the Method of Simulated Moments~\cite{Gourieroux-Monfort-97}, and the Two-Stage (TS) Approach~\cite{garatti2013new,garatti2008estimation}.

Out of these new methods, TS is an attractive alternative to traditional parametric techniques, since it only requires one to be able to simulate data for a fixed choice of the parameters, and, by properly choosing its second stage, it requires only solving convex optimization problems (thus avoiding getting trapped into local optima). The TS approach works by seeing the estimation problem as an inverse \emph{supervised learning} setup~\cite{Shalev-Shwartz-Ben-David-12}, where one simulates a large number of samples for different values of the parameters, and uses any supervised learning method to predict the values of the parameters for a given data sample. The supervised learning problem is decomposed into two stages, where the first stage involves ``compressing'' the data into a small set of values, and the second one uses such compressed samples to ``predict'' the values of the parameters. In spite of its advantages with respect to other approaches, however, TS relies on several user choices, whose impact on the performance of the method is not yet clear.

In this paper we provide a formalization of TS, by deriving it within a statistical decision theory setup. This allows us to note that the standard TS is actually a Bayes estimator (with respect to a given prior on the parameters), and to extend the approach to minimax decision rules. In addition, we focus on the case of independent and identically distributed (\emph{i.i.d.}) data, for which we suggest a specific choice for its first (or ``compression'') and second stages, which leads to convex optimization problems that can be efficiently solved using off-the-shelf solvers.

In summary, the main contributions of this paper are:
\begin{itemize}
\item We provide a statistical decision-theoretical derivation of TS, which leads to Bayes and minimax formulations, and give a reason for decomposing the problem into two stages;

\item we suggest a specific structure for the second stage of TS, which leads to simple convex programs for both Bayes and minimax formulations;

\item for \emph{i.i.d.} data, we propose a specific choice for the first stage of TS, based on sample quantiles of the data; and

\item we illustrate the performance of the novel Bayes and minimax TS formulations with a numerical example based on a Weibull distribution.

%\item \BL{-Need to write this precisely} Build simpler models in the second stage of TS approach that are interpretable  
\end{itemize}

The rest of this work is organized as follows. In Section~\ref{sec: problem formulation}, the statistical decision theory setup for parameter estimation is presented. Section~\ref{sec:SDT} describes a computational way to approximate Bayes and minimax estimators, and explains their connection to TS. In Section~\ref{sec:IID_case}, the Bayes and minimax formulations of TS are specialized for \emph{i.i.d.} data. A simulation study assessing the performance of the above formulations is presented in Section~\ref{sec: simulations}, and finally conclusions are drawn in Section~\ref{sec: conclusions}.

\section{Problem Formulation} \l{sec: problem formulation}

Consider an environment, or data generating mechanism, that produces a quantity $\mathbf{y} \in \mathcal{Y}$ according to a probability distribution $\P(\mathbf{y}|\t)$ determined by an unknown parameter $\theta \in \Theta \subseteq \mathbb{R}^d$, where $d \geq 1$.
Based on the value of $\mathbf{y}$, the goal is to design an \emph{estimator} (or \emph{decision rule}) $\delta\colon \mathcal{Y} \to \Theta$ of the true unknown parameter $\t$ such that the \emph{risk}  \begin{align}
R(\t,\d) := \E_{\mathbf{y} \sim \P(\mathbf{y}|\t)} \{L(\t,\d(\mathbf{y}))\} \label{8}
\end{align}
is minimized, where $L\colon \Theta \times \Theta \to \R_0^+$ is a \emph{loss} function. In other words, $L(\t,\delta(\mathbf{y}))$ is the loss of estimating the parameter as $\delta(\mathbf{y})$ when its true value is $\t$. Since the risk depends on $\t$, to evaluate the performance of estimators $\delta(\mathbf{y})$, it is required to reduce the risk to a function that depends only on $\delta$. The notion of an ``optimal" decision rule is then defined with respect to this function. In this paper, we consider the following types of optimal rules:

\begin{enumerate}
\item \textbf{Bayes rules}: Given a probability distribution $\P(\t)$ on $\Theta$, a decision rule $\delta^*_\text{Bayes}$ is said to be a \emph{Bayes rule} (with respect to $\P(\t)$) if it minimizes $\E_{\t \sim \P(\t)}[R(\t, \delta)]$.

\item \textbf{Minimax rules}: A decision rule $\d^{*}_{\text{minimax}}$ is called a \emph{minimax rule} if it minimizes $\max\nolimits_{\t \in \Theta} R(\t, \delta)$.
\end{enumerate}

In some of the subsequent sections, we will consider the case where $\mathbf{y}=(y^1,\ldots,y^N)$, \ie, $\mathbf{y}$ is a vector consisting of multiple observations.

\section{A Statistical Decision-Theoretical Formulation of TS}\label{sec:SDT}

In this section, we will show how to derive, in a computational manner, Bayesian and minimax decision rules, and describe their connection to the Two-Stage Approach.

\subsection{Bayes estimator} \label{subsec:Bayes}
Expanding the expression for $R(\t,\d)$ in \eqref{8}, we have
\begin{align*}
R(\t,\d) &= \E_{\mathbf{y} \sim \P_{(\mathbf{y}|\t)}} \lrs{L\lrp{\t,\d(\mathbf{y})}} = \i{\mathcal{Y}}{} L\lrp{\t,\d(\mathbf{y})} \P(\mathbf{y}|\t) \, d\mathbf{y}.
\end{align*}
Now, we define \emph{weighted risk}
\begin{align}
R_{\text{weighted}}(\d) = \i{\Theta}{} R(\t,\d) \pi(\t) \, d\t,
\end{align}
where $\pi\colon \Theta \to \mathbb{R}_0^+$ is a weighting function over $\Theta$. Note that $\pi$ need not be a probability density function (\emph{pdf}) over $\Theta$, that is, $\int_{\Theta} \pi(\t) d\t$ does not need to be equal to $1$ (or even finite). However, if $\pi(\t) = \P(\t)$ happens to be a pdf over $\mathbb{R}^d$, then we obtain the definition of \emph{Bayes risk},
\begin{align} \label{eq:Rbayes}
R_{\text{Bayes}}(\d) = \i{\Theta}{} R(\t,\d) \P(\t) \, d\t,
\end{align}
where $\P(\t)$ is called a \emph{prior} on $\t$. The minimizer of $R_{\text{Bayes}}$ is called the \emph{Bayes rule} associated with $\P(\t)$.

The standard approach to derive the Bayes rule is as follows. We first expand \eqref{eq:Rbayes}:
\begin{align}
R_{\text{Bayes}}(\d) &= \i{\Theta}{} \i{\mathcal{Y}}{} L\lrp{\t,\d(\mathbf{y})} \P(\mathbf{y}|\t) \P(\t) \, d\mathbf{y} d\t \nn \\
%&= \i{\mathcal{Y}}{} \i{\mathbb{R}^d}{} L\lrp{\t,\d(\mathbf{y})} \P(\t|\mathbf{y}) \P(\mathbf{y}) \, d\t d\mathbf{y} \nn \\
&=\i{\mathcal{Y}}{} \lrs{\i{\Theta}{} L\lrp{\t,\d(\mathbf{y})} \P(\t|\mathbf{y}) \, d\t} \P(\mathbf{y}) \, d\mathbf{y}. \l{12}
\end{align}
In \eqref{12}, $\P(\mathbf{y})$ is the \emph{evidence} or \emph{marginal likelihood} of the data $\mathbf{y}$, and is given by $\P(\mathbf{y}) := \i{\Theta}{} \P(\mathbf{y}|\t) \P(\t) d\t,$
%
% \begin{align*}
% \P(\mathbf{y}) := \i{\Theta}{} \P(\mathbf{y}|\t) \P(\t) d\t,
% \end{align*}
%
while $\P(\t|\mathbf{y})$ is the \emph{posterior distribution} of $\t$ given the data $\mathbf{y}$, defined as $\P(\t|\mathbf{y}) %
:= \frac{\P(\mathbf{y}|\t) \P(\t)}{\P(\mathbf{y})} %
= \frac{\P(\mathbf{y}|\t) \P(\t)}{\i{\Theta}{} \P(\mathbf{y}|\tilde{\t}) \P(\tilde{\t}) d\tilde{\t}}$. 
%
% \begin{align*}
% \P(\t|\mathbf{y}) %
% := \frac{\P(\mathbf{y}|\t) \P(\t)}{\P(\mathbf{y})} %
% = \frac{\P(\mathbf{y}|\t) \P(\t)}{\i{\Theta}{} \P(\mathbf{y}|\tilde{\t}) \P(\tilde{\t}) d\tilde{\t}}.
% \end{align*}
%
It is clear from \eqref{12} that, in order to minimize $R_{\text{Bayes}}$ we have to minimize the inner integral $\i{\Theta}{} L(\t,\d(\mathbf{y})) \P(\t|\mathbf{y}) d\t$ over $\d$, since $\P(\mathbf{y})$ is non-negative.
Therefore, the Bayes decision rule is given by
$\d^{*}_{\text{Bayes}} = \text{arg} \, \min_{\d} \, \i{\Theta}{} L(\t,\d(\mathbf{y})) \P(\t|\mathbf{y}) d\t.$

We now propose an alternative computational approach to derive the Bayes rule associated with a prior $\P(\t)$, which essentially corresponds to the TS method~\cite{garatti2008estimation,garatti2013new}. To this end, instead of using the last equation in \eqref{12}, our starting point is the first equation in \eqref{12}, which we write as
\begin{align}
R_{\text{Bayes}}(\d)
%&= \i{\Theta}{} \i{\mathcal{Y}}{} L\lrp{\t,\d(\mathbf{y})} \P(\mathbf{y}|\t) \P(\t) \, d\mathbf{y} d\t \nn \\
&= \i{\Theta}{} \lrs{\i{\mathcal{Y}}{} L(\t,\d(\mathbf{y})) \, \P(\mathbf{y}|\t) d\mathbf{y}} \P(\t) d\t. \label{14}
\end{align}
Now, consider the inner integral of \eqref{14}, which is a function of $\t$, since it integrates out $\mathbf{y}$. Let us define this integral as
\begin{align}
f(\t) := \i{\mathcal{Y}}{} L(\t,\d(\mathbf{y})) \, \P(\mathbf{y}|\t) d\mathbf{y}.  \l{15}
\end{align}
We can approximate $f(\t)$ using Monte-Carlo sampling \cite{P-766},
\begin{align}
    f(\t) \approx \dfrac{1}{M_\mathbf{y}} \S{j=1}{}{M_\mathbf{y}} L(\t,\d(\mathbf{y}_j)), \l{16}
\end{align}
where $\mathbf{y}_1, \dots, \mathbf{y}_{M_\mathbf{y}}$ are independent random samples from $\P(\mathbf{y}|\t)$.
Substituting \eqref{16} in \eqref{14} yields
\begin{align}
R_{\text{Bayes}}(\d) &\approx \i{\Theta}{} \lrs{\dfrac{1}{M_\mathbf{y}} \S{j=1}{}{M_\mathbf{y}} L(\t,\d(\mathbf{y}_j))} \P(\t) \, d\t. \l{17}
\end{align}
We can further approximate the outer integral of \eqref{17} by Monte-Carlo sampling, by taking $M_\t$ independent samples $\t_1, \dots, \t_{M_\t}$ from $\P(\t)$:
\begin{align}
    R_{\text{Bayes}}(\d) &\approx \dfrac{1}{M_\mathbf{y}} \dfrac{1}{M_\t} \S{i=1}{}{M_\t} \S{j=1}{}{M_{\mathbf{y}}} L(\t_i,\d(\mathbf{y}_{ij})), \l{18}
\end{align}
where $\{\mathbf{y}_{ij}\} \stackrel{i.i.d.}{\sim} \P(\mathbf{y}|\t_i)$ and $\{\t_i\} \stackrel{i.i.d.}{\sim} \P(\t)$, for $i = 1,\ldots,M_\t$ and $j = 1,\ldots,M_\mathbf{y}$.

From \eqref{18}, it follows that one can approximate the Bayes rule by solving the optimization problem
\begin{align} \label{eq:approx_Bayes}
\tilde{\d}_{\text{Bayes}} := \arg \min_{\d \in \Delta} \dfrac{1}{M_\mathbf{y}} \dfrac{1}{M_\t} \S{i=1}{}{M_\t} \S{j=1}{}{M_\mathbf{y}} L(\t_i,\d(\mathbf{y}_{ij})),
\end{align}
where $\Delta$ is a suitable function space. If $M_\mathbf{y}$, $M_\t$ are sufficiently large, and $\Delta$ is sufficiently large, one should expect that $\tilde{\d}_{\text{Bayes}} \approx \d^*_{\text{Bayes}}$ in a suitable sense.

\subsection*{Choosing the function space}
The next step is to choose a function space $\Delta$. To this end, notice that typically $\mathcal{Y}$ is a product space, \ie, $\mathcal{Y} = \tilde{\mathcal{Y}}^N$, so that $\mathbf{y} \in \mathcal{Y}$ is of the form $\mathbf{y} = (y^1, \dots, y^N)^T$, where $y^i \in \tilde{\mathcal{Y}}$ ($i = 1, \dots, N$) are the individual \emph{samples} in the data, and $N$ is the \emph{sample size}.
Furthermore, $\P(\mathbf{y}|\t)$ may also have a product structure, \ie, if $(y^1, \dots, y^N)^T \sim \P(\mathbf{y}|\t)$, then $y^i \stackrel{i.i.d.}{\sim} \P(\tilde{y}|\t)$, or $\P(\mathbf{y}|\t)$ may correspond to a segment of a stationary distribution; in those cases, the measure $\P(\mathbf{y}|\t)$ may be subject to a measure concentration phenomenon~\cite{ledoux2001}, which means that $\P(\mathbf{y}|\t)$ may be highly concentrated on a subset of $\mathcal{Y}$. This implies that the decision rule $\d^*_{\text{Bayes}}$ may be difficult to approximate using standard functional approximation methods.

To overcome this issue, we will add the subscript $N$ to $\d$, to emphasize the dependence of the number of samples $N$, and express $\d_N$ as the composition of two functions,
\begin{align} \l{eq:decomp_delta}
\d_N = g_N \circ h_N,
\end{align}
where $h_N\colon \mathcal{Y} \to \mathbb{R}^n$ and $g_N\colon \mathbb{R}^n \to \Theta$, with $n \ll N$. For identifiability purposes~\cite{LehmCase98}, we require that $N$ is larger than $n$. 
Function $h$ can be chosen as a fixed function (\ie, not to be optimized with respect to) with the property that if $\mathbf{y} \sim \P(\mathbf{y}|\t)$, then $h_N(\mathbf{y}) \to \bar{h}(\mathbf{y})$ in probability as $N \to \infty$, where $\bar{h}\colon \mathcal{Y} \to \mathbb{R}^n$ is differentiable and does not depend on $N$. This means that the measure concentration of $\P(\mathbf{y}|\t)$ as $N$ grows can be taken care of by $h_N$, and if $N$ is sufficiently large, $g_N$ can be designed independently of $N$, \emph{i.e.}, $g_N \equiv g$ for a fixed function $g\colon \mathbb{R}^n \to \Theta$.

A possibility for $h_N$ is to use an estimator for a \emph{misspecified model} of $\mathbf{y}$, \ie, a model which is not related to $\P(\mathbf{y}|\t)$, but which is nevertheless easy to estimate (using, for instance, least squares~\cite{casella2021statistical}). The value of $h_N(\mathbf{y})$ should serve as a \emph{approximate sufficient statistic}~\cite{approximatesuff} for $\t$, in the sense of containing, approximately, the same information about $\t$ as $\mathbf{y}$. Notice also that this resembles the use of misspecified models in techniques such as Indirect Inference~\cite{Gourieroux}.

Once $h_N$ is chosen, we are left with the problem of determining $g$ according to
\begin{align}\l{eq:Bayesian_obj}
g^* = \arg \min_{g \in G} \dfrac{1}{M_\mathbf{y}} \dfrac{1}{M_\t} \S{i=1}{}{M_\t} \S{j=1}{}{M_\mathbf{y}} L(\t_i,g(h_N(\mathbf{y}_{ij}))).
\end{align}
Here, $G$ is a function space that can be chosen according to general functional approximation methods, such as deep neural networks, boosted decision trees, radial basis functions, or finite dimensional function spaces (\eg, polynomials, piecewise affine functions, \dots).

\begin{remark} \label{rem:stationarity}
Under i.i.d. or stationarity assumptions, if $N$ is sufficiently large, each sample will give a good estimate of the loss (due to a concentration of measure), so we can take $M_\mathbf{y} = 1$ without significant loss in performance.
\end{remark}

The approach given here to approximate $\d^*_\text{Bayes}$ is essentially identical to the TS approach given in ~\cite{garatti2013new}, where the first stage consists in the computation of $h_N(\mathbf{y}_{ij})$ (called the ``compression stage''), and in the second stage $g^*$ is determined. However, our formulation clarifies that the way how the values of $\t$ are sampled (via $\P(\t)$) affects the resulting estimate, which is in fact an approximate Bayes estimator; for this reason, the estimator $\d^*_N$ will be called the \emph{Bayes TS estimator} of $\t$ under the prior $\P(\t)$.

% In the next subsection, we consider a special case of posterior mean estimation of a single parameter, where $M_D = 1$, $G$ consists of all linear functions $g\colon \mathbb{R}^n \to \A=\mathbb{R}$ and $L$ is the MSE loss.

\subsection{Minimax estimator} \label{subsec:minimax}
In this subsection we modify the previous formulation in order to arrive at a minimax estimator. Let us again recall \eqref{8}. The minimax estimator is given by $\d^*_{\text{minimax}} = \text{arg} \, \min_{\d \in \Delta} \left[\max_{\t \in \Theta} \, R(\t,\d) \right].$

%Equivalently, the minimax rule is obtained by the following optimization problem:
%%
%\begin{align}\l{eq:minimax_opt}
%    \min_{\d \in \Delta} \, \max_{\t \in \mathbb{R}^d} \, R(\t,\d).
%\end{align}
%%
Let us rewrite the inner optimization problem $\max_{\t \in \Theta} R(\t,\d)$ as
\begin{align}\l{eq:max_pi}
\max_{\t \in \Theta} \, R(\t,\d) = \max_{\pi \in \mathcal{P}(\Theta)} \i{\T}{} R(\t,\d) \, \pi(\t) \, d\t,
\end{align}
where $\mathcal{P}(\Theta)$ is the space of probability distributions in $\Theta$, and $\pi(\cdot)$ can be interpreted as a prior distribution on $\t$. 

By substituting the definition of \emph{risk} $R(\t,\d)$ in \eqref{eq:max_pi}, we have
\begin{align}\l{eq:max_pi_fr}
\max_{\t \in \Theta} \, R(\t,\d) = \max_{\pi \in \mathcal{P}(\Theta)}  \i{\Theta}{}  \i{\mathcal{Y}}{} L\lrp{\t,\d(\mathbf{y})} \P(\mathbf{y}|\t) \, \pi(\t) \, d\mathbf{y} d\t.
\end{align}
Using\eqref{eq:max_pi_fr}, we obtain% an equivalent characterization of $\d^*_{\text{minimax}}$:
\begin{align}\l{eq:equi_opt}
\d^*_{\text{minimax}} &= \min_{\d \in \Delta} \max_{\pi \in \mathcal{P}(\Theta)} \i{\Theta}{}  \i{\mathcal{Y}}{} L\lrp{\t,\d(\mathbf{y})} \P(\mathbf{y}|\t) \, \pi(\t) \, d\mathbf{y} d\t \nn \\
&= \min_{\d \in \Delta} \, \max_{\pi \in \mathcal{P}(\Theta)} \i{\Theta}{}  \i{\mathcal{Y}}{} L\lrp{\t,\d(\mathbf{y})} \P(\mathbf{y}|\t) \, d\mathbf{y} \pi(\t) \, d\t,
\end{align}
assuming that we can interchange the integrals in \eqref{eq:equi_opt}. Now, we can approximate the inner integral using Monte-Carlo sampling as follows:
\begin{align}\l{eq:first_appx}
\i{\mathcal{Y}}{} L\lrp{\t,\d(\mathbf{y})} \P(\mathbf{y}|\t) \, d\mathbf{y} \approx \dfrac{1}{M_\mathbf{y}} \S{j=1}{}{M_\mathbf{y}} L(\t,\d(\mathbf{y}_{j,\t})),
\end{align}
where $\{\mathbf{y}_{j,\t}\} \stackrel{i.i.d.}{\sim}{\P(\mathbf{y}|\t)}$ for a fixed $\t$, and $M_\mathbf{y}$ is the number of samples generated according to $\P(\mathbf{y}|\t)$. 

We now use \eqref{eq:first_appx} in \eqref{eq:equi_opt} to obtain 
\begin{multline}
\min_{\d \in \Delta} \max_{\pi \in \mathcal{P}(\Theta)} \i{\Theta}{} \i{\mathcal{Y}}{} L\lrp{\t,\d(\mathbf{y})} \P(\mathbf{y}|\t) \, \pi(\t) \, d\mathbf{y} d\t \\
\approx \min_{\d \in \Delta} \, \max_{\pi \in \mathcal{P}(\Theta)} \i{\Theta}{} \lrs{\dfrac{1}{M_\mathbf{y}} \S{j=1}{}{M_\mathbf{y}} L(\t,\d(\mathbf{y}_{j,\t}))} \pi(\t) \, d\t.
\end{multline}
Therefore, the new and approximate optimization problem will be posed as 
\begin{align}\l{eq:new_min_max}
\min_{\d \in \Delta} \, \max_{\pi \in \mathcal{P}(\Theta)} \i{\Theta}{} \lrs{\dfrac{1}{M_\mathbf{y}} \S{j=1}{}{M_\mathbf{y}} L(\t,\d(\mathbf{y}_{j,\t}))} \pi(\t) \, d\t.
\end{align}

In contrast to Bayes rules, we will now apply \emph{importance sampling} \cite{mcbook} to further approximate the integral in \eqref{eq:new_min_max}. Notice that the optimization problem in \eqref{eq:new_min_max} involves a maximization over $\pi \in \mathcal{P}(\Theta)$. If we used Monte-Carlo sampling instead, then we would need to fix $\pi$, and for each choice of $\pi$ we would need to generate samples of $\t$ according to $\pi$; finally, we would have to find the value of $\pi$ for which the approximate Monte-Carlo sum is a maximum for a fixed choice of $\d$. This is computationally tedious. An alternative approach, through Importance Sampling, is to fix a distribution $s(\t)$, called a \emph{proposal distribution} (to be decided by a user) from which we can sample values of $\t$. We use these samples to approximate the integral in \eqref{eq:new_min_max} by employing \emph{self-normalized importance sampling}\cite[Chapter 9, page 8]{mcbook} as
\begin{align}\l{eq:min_max_is}
\min_{\d \in \Delta} \, \max_{w_1, \dots, w_{M_\t} \in \mathbb{R}} \, &\dfrac{1}{M_\mathbf{y}} \dfrac{1}{M_{\t}} \S{j=1}{}{M_\mathbf{y}} \S{i=1}{}{M_{\t}} L({\t}_i,\d(\mathbf{y}_{ij})) w_i \nn \\
\text{s.t} \qquad &w_i  \geq 0, \quad i=1, \dots, M_\t  \\
& \sum_{i=1}^{M_\t} w_i = 1, \nn
\end{align}
where $\mathbf{y}_{ij} = \mathbf{y}_{j, \t_i}$, $w_i = \dfrac{\pi(\t_i)/s(\t_i)}{\sum_{k=1}^{M_\theta} \pi(\t_k)/s(\t_k)}$, and $\{\t_i\} \stackrel{i.i.d.}{\sim} s(\t)$.

% One should be careful when choosing $q$. The support of the distribution $q$ should contain the support of all possible distributions $\pi$ (that is, $\mathbb{R}^d$) and the tail of the distribution $q$ should be as heavy as the tail of the optimal $\pi$, say, $\pi^*$. 

%The optimal choice of $q$, from a variance point of view, is given by $q(\t) \propto \abs{\S{j=1}{}{M_\mathbf{y}}L(\t,\d(\mathbf{y}_{i,\t}))} \pi^*(\t)$; this quantity is, unfortunately, unknown since we do not know $\pi^*$ in advance.

% By using importance sampling, notice that now we do not need to sample values of $\t$ for every possible choice of $\pi$, and that \eqref{eq:min_max_is} can be solved analytically.

% \CR{- Not sure that \eqref{eq:min_max_is} can always be solved analytically: this depends on $L$ and the parameterization of $\d$.}

% \BL{- Shouldn't the sum over $w_i$ be ``approximately" equal to 1 ? } 

% \BL{ - Need to give a strong motivation as to why we resort to Importance Sampling in the second step. It can be thought of variance reduction method as opposed to Monte-Carlo sampling, but one possible question that we might ask: why we are not applying importance sampling in \eqref{eq:first_appx}. A possible explanation to that would be the fact that we might not know the distribution structure of $\P(D|\t)$. This might be the case if we know the model perfectly, that is, the exact dependence of input-output pair (in the example, just the output alone) }

%\subsection{Solving the Minimax Formulation}

\medskip
Now we will simplify the minimax optimization problem \eqref{eq:min_max_is}.
%By ignoring the constants $M_{\t},M_{\mathbf{y}}$ that appear in the denominator, we obtain an equivalent minimax problem given by 
%%
%\begin{align}\l{eq:min_max_is_2}
%\min_{\d \in \Delta} \, \max_{w_1, \dots, w_{M_\t} \in %\mathbb{R}} & \S{j=1}{}{M_\mathbf{y}} \S{i=1}{}{M_{\t}} L({\t}_i,\d(\mathbf{y}_{ij})) w_i \nn \\
%\text{s.t.} \qquad & w_i  \geq 0, \quad i=1, \dots, M_\t \\
%& \sum_{i=1}^{M_\t} w_i = 1. \nn
%\end{align}
%%
%We can interchange the order of summation in \eqref{eq:min_max_is_2} and obtain the equivalent problem
%
%\begin{align}\l{eq:min_max_is_3}
%\min_{\d \in \Delta} \, \max_{w_1, \dots, w_{M_\t} \in %\mathbb{R}} & \S{i=1}{}{M_{\t}} \S{j=1}{}{M_\mathbf{y}}L({\t}_i,\d(\mathbf{y}_{ij})) w_i \nn \\
%\text{s.t.} \qquad & w_i  \geq 0, \quad i=1, \dots, M_\t \\
%& \sum_{i=1}^{M_\t} w_i = 1. \nn
%\end{align}
%%
%Now let us define $L_i(\d) := \S{j=1}{}{M_D}L({\t}_i,\d(\mathbf{y}_{ij}))$ and rewrite \eqref{eq:min_max_is_3} as follows:
By ignoring the constants $M_{\t},M_{\mathbf{y}}$ that appear in the denominator, and defining $L_i(\d) := \S{j=1}{}{M_\mathbf{y}}L({\t}_i,\d(\mathbf{y}_{ij}))$, we can write \eqref{eq:min_max_is} as
\begin{align}\l{eq:min_max_is_reduced}
\min\limits_{\d \in \Delta} \, \max\limits_{w_1, \dots, w_{M_\t} \in \mathbb{R}}  & \S{i=1}{}{M_{\t}} L_i(\d) w_i \nn \\
\text{s.t.} \qquad &w_i  \geq 0, \quad i=1, \dots, M_\t \\
& \sum_{i=1}^{M_\t} w_i = 1. \nn
\end{align}
Furthermore, we can notice that the inner maximization problem takes place over a probability simplex, so the maximum is obtained by assigning probability mass $1$ to $\max_{i = 1, \dots, M_{\t}} L_i(\d)$. Therefore, by using the epigraph formulation of the optimization problem~\cite[page~134]{boyd_vandenberghe_2004}, one can show that \eqref{eq:min_max_is} is equivalent to
\begin{align}\l{eq:min_max_cvx}
\begin{array}{cl}
\displaystyle \min_{\d \in \Delta, t \in \R} & t  \\
\text{s.t.} & L_i(\d) \leq t, \quad i = 1, \dots, M_{\t}.
\end{array}
\end{align}

The form of the decision rule $\d$ can be taken, as in the Bayesian case, as $\d = g \circ h_N$ (\emph{cf.} \eqref{eq:decomp_delta}). Due to its similarity to TS, we will call it the \emph{minimax TS estimator}.

\medskip
Finally, we will consider a very special case of the minimax TS estimator, where $d = 1$, the loss function is the \emph{mean square error} $L(\t, \hat{\t}) = (\t - \hat{\t})^2$, where $\hat{\t}$ is an estimate of $\t$ and $g$ is a linear function\footnote{This special case can be easily extended to more than one parameter, \ie, $d > 1$, by applying TS to estimate each parameter separately.}; furthermore, we assume that $\mathbf{y}$ is a vector of \emph{i.i.d.} or stationarity entries, so by Remark~\ref{rem:stationarity} we take $M_{\mathbf{y}}=1$. These restrictions will allow us to derive an explicit computation procedure for the second stage. In this case, $L_i(\d) = L({\t}_i,\d(\mathbf{y}_{i1}))$. Furthermore, $\d(\mathbf{y}_{i1})=\beta^T\alpha_i$, where the vector $\alpha_i \in \mathbb{R}^n$ contains the compressed data obtained in the first stage of TS, and $\beta \in \mathbb{R}^n$ is the linear regression coefficient vector that describes the linear function in the second stage of TS. Then, we have that
\begin{align}\l{eq:loss_func_cvx}
L(\t_i,\d(\mathbf{y}_{ij})) &= \lrp{\t_i-\beta^T\alpha_i}^2 \nn \\
%&=\beta^T\lrp{\alpha_i\alpha_i^T}\beta + \lrp{-2\t_i\alpha_i}^T\beta+\t_i^2 \nn \\
&= \beta^T\lrp{\alpha_i\alpha_i^T}\beta -2 \theta_i \alpha_i^T\beta + \theta_i^2.
\end{align}
Substituting \eqref{eq:loss_func_cvx} in \eqref{eq:min_max_cvx} we obtain
\begin{align}\l{eq:min_max_cvx_1}
\begin{array}{cl}
\displaystyle \min_{\beta \in \mathbb{R}^n, t \in \R} & t  \\
\text{s.t.} &\beta^T\lrp{\alpha_i\alpha_i^T}\beta - 2 \theta_i \alpha_i^T\beta + \theta_i^2 \leq t,\\
&\qquad\qquad\qquad\qquad\quad i = 1, \dots, M_{\t}.
\end{array}
\end{align}
%
% We can equivalently express this optimization problem as a semidefinite program by using the Schur complement formula \cite{boyd1}:
%
% \begin{align} \label{eq:min_max_schur}
% \begin{array}{cl}
% \displaystyle \min_{\beta \in \mathbb{R}^n, t \in \R} & t  \\
% \text{s.t.} &\begin{pmatrix} t+2\t_i\alpha_i^T\beta-\t_i^2 & \beta^T\alpha_i \\ \beta^T\alpha_i & 1 \end{pmatrix} \succeq \mathbf{0},\ i = 1, \dots, M_{\t}.
% \end{array}
% \end{align}
%
This optimization problem can be solved using a standard convex programming solver, such as CVXPY~\cite{diamond2016cvxpy}.

\section{Two-Stage Approach for i.i.d. Data} \l{sec:IID_case}

The formulations of TS given in Section~\ref{sec:SDT} make no special assumptions on the distribution of the data $\mathbf{y}$. In this section, we assume that $\mathbf{y}$ is a vector whose entries, $\{ y^i \}$, are \emph{i.i.d.} according to distribution $f(\t)$, where $f(\t)$ is a \emph{pdf} parameterized by $\t \in \Theta$.

To properly account for the \emph{i.i.d.} assumption, we notice that under it, reasonable decision rules are ``permutation-invariant'' \cite{LehmCase98}, that is, their output should not vary if the entries of $\mathbf{y}$ are permuted. Such decision rules can be shown~\cite{LehmCase98} to be functions of the \emph{order statistics} of $y^1, \dots, y^N$, which correspond to the values of $y^1,\ldots,y^N$ sorted in ascending order, and are denoted by $y^{(1)}, \dots, y^{(N)}$.

In the original formulation of TS~\cite{garatti2013new, garatti2008estimation}, the coefficients $\alpha_1,\ldots \alpha_n$ of an estimated $\text{AR}(n)$ model (based on $\mathbf{y}$) are used in the first stage of the method to compress the samples $y^1,\ldots,y^N$, where $n \ll N$. This is a reasonable way to compress the observations when they are realizations of a non-stationary process. However, if the data is \emph{i.i.d.}, then we suggest to compress the data starting from the order statistics. To this end, we will use some \emph{quantiles} of the order statistics: If $0 < p < 1$, then the $(100 p)$-th sample quantile is a linear interpolation between $y^{(Np)}$ and $y^{(N(1-p))}$. The computation of a small and fixed number of quantiles as the compressed data constitutes the first stage of TS, and the description of the first stage is given below.

\subsection{First Stage}
For the first stage, we set $p = k / n$, where $k=1, \dots, n$, and obtain $n$ quantiles. We then collect those quantiles into the compressed $n$-dimensional real-valued column vector $\alpha = (\alpha_1,\ldots,\alpha_n)^T$.  The first stage for both Bayes TS and minimax TS is the same.

\subsection{Second stage}
For both formulations, the second stage starts by constructing non-linear features $\phi(\alpha)$ using the compressed data $\alpha$ and a fixed non-linear function $\phi\colon \R^n \ra \R^m$ with $n < m < N$.
The function $\phi$ must be chosen by the user, and it depends on the type of parameter(s) that we want to estimate. In both formulations, we then design $g$ as a linear function in composition with $\phi$, \ie, as $g(\alpha) = \beta^T \phi(\alpha)$, where $\beta$ is determined as explained in Section~\ref{sec:SDT}. If $\t=(\t_1,\ldots,\t_d)^T$, where $d>1$, is the parameter vector to be estimated, then we can design a separate estimator $g_i(\alpha)={\beta_i}^T\alpha$ for each $i = 1, \dots, d$.
%estimates for $\t_1,\ldots,\t_d$ can be determined by designing $g_1(\alpha)={\beta_1}^T\alpha,\ldots,g_d(\alpha) = {\beta_d}^T\alpha$ for $\t_1,\ldots,\t_d$ respectively.}
To avoid numerical issues, in both Bayes and minimax formulations we add a regularization term of the form $\lambda \| \beta \|_2^2$ to the optimization problem to be solved, where $\lambda > 0$ is a very small number (\eg, $10^{-8}$).

%For the second stage, there are two subtle differences between the Bayes and minimax formulations. First, a prior distribution on $\t$ is assumed in the Bayes setup, whereas no such assumption is made in the minimax formulation, even though a proposal distribution $q$ needs to be chosen. Second, the objective functions that need to be minimized in the Bayes and minimax formulations are different (cf. \eqref{eq:Bayesian_obj} and \eqref{eq:min_max_cvx_1}).

%In the minimax formulation, we consider a linear function $g$ which is determined as presented in Section~\ref{subsec:minimax}.
%In the Bayes setup, instead, kernel ridge regression is deployed. We construct non-linear features $\phi(\alpha)$ using the the compressed data $\alpha$, where $\phi$ is non linear function such that $\phi : \R^{n} \ra \R^{m}$ with $n<m<N$. These non-linear features $\phi = (\phi_1,\ldots,\phi_m)^T$ are then fitted against the true values of parameter $\t$. The choice of non-linear function $\phi$ depends on the type of parameter that we want to estimate. 

\section{Simulations} \l{sec: simulations}
In this section, we illustrate the performance of the Bayes and minimax estimators derived in Section~\ref{sec:SDT} on the estimation of the two parameters of a Weibull distribution. Weibull distributions are widely used in reliability engineering\cite{weibull1,weibull2} to model the probability of failure of an equipment at a particular age, and are parameterized by: (i) a \emph{scale} parameter, denoted by $\eta > 0$, and (ii) a \emph{shape} parameter, denoted by $\gamma > 0$. Here, $\eta$ and $\gamma$ are unknown. Hence, this is a two dimensional parameter estimation problem, with $\t = (\eta, \gamma)^T$ being the unknown parameter. As described in Section~\ref{sec:IID_case}-B, we design separate estimators for $\eta$ and $\gamma$, which are denoted by $\h{\eta}$ and $\h{\gamma}$ respectively.

% The Weibull distribution is a continuous probability distribution widely used in the field of reliability engineering.
The Weibull probability density function is
%The Weibull cumulative distribution function (\emph{c.d.f.}) describing the age of an equipment is
%
\begin{align}
f(A) &= \frac{\gamma}{\eta} \lrp{\frac{A}{\eta}}^{\gamma -1} \exp\!\left[-\lrp{\frac{A}{\eta}}^{\gamma}\right], \quad A \geq 0.
\end{align}
% %
%
%
%\begin{align}\l{eq:weibull}
%F(A) &= 1 - \exp\left[-\lrp{\frac{A}{\eta}}^\gamma\right], \quad A \geq 0.
%\end{align}
%
%where $A$ in this application is the equipment age.

We follow the TS approach described in Section~\ref{sec:IID_case}, where for the first stage we consider $n = 10$ quantiles, and for the second stage, we choose $\phi(\alpha)$ according to the following:
\begin{itemize}
\item For scale parameter $\eta$, 
\begin{align}
\phi_i(\alpha) &= \begin{cases}
\alpha_i, & \text{if } 1 \leq i \leq n \\
\displaystyle \frac{\alpha_{i-n+1}}{\alpha_1}, & \text{if } n+1 \leq i \leq 2n-1.
\end{cases}
\end{align}
The reason for choosing these combinations of quantiles to estimate $\eta$ is that quotients of the (theoretical) quantiles of a distribution are proportional to its scale parameter (when the other parameters are kept fixed), which suggests that the quantities $\phi_i(\alpha)$ above can be used for estimating $\eta$.

\item For shape parameter $\gamma$, let
\begin{align}
\psi_j(\alpha) &= \begin{cases}
\alpha_j, & \text{if } 1 \leq j \leq n \\
\displaystyle \frac{\alpha_{j-n}}{\alpha_n}, & \text{if } n+1 \leq j \leq 2n-1.
\end{cases}
\end{align}
We let $\phi(\alpha)$ consist of monomials of the $\psi_j(\alpha)$'s up to order $2$. This choice is suggested by the fact that a small increase in $\gamma$ (for $\gamma > 1$) leads to a similar change of the Weibull distribution as a small decrease in $\eta$.
\end{itemize}

We should remark that for the Weibull distribution there exist exact expressions for $\eta$ and $\gamma$ in terms of any two quantiles~\cite{johncook}. However, using such expressions together with sample quantiles may lead to sub-optimal estimates, since they ignore the variability of the sample quantiles.

% \CR{- A reference to the above statement is here: \url{https://www.johndcook.com/quantiles_parameters.pdf}. Can you please add it to the .bib file?}

%\BL{- Scale parameter $\eta$ stretches the \emph{pdf} while $\gamma$ is kept constant. That is, the distribution gets stretched out to the right and its height decreases if $\eta$ is increased. Similarly, the distribution gets pushed in towards the left and its height increases if $\eta$ is decreased. Hence, in order to capture the amount of stretch in \emph{pdf} from the order statistics, we divide the ordered statistics $\alpha_2=y^{(2)},\ldots,\alpha_n = y^{(n)}$ by $\alpha_1 = y^{(1)}$ to get representative features for the scale parameter $\eta$.}

We will now look at results corresponding to both Bayes and minimax frameworks that we discussed in Section~\ref{sec:SDT}.

%Specifically, we show the scatter plots of estimated values of $\t$ vs. true values, for different true values of $\t$. 

%Since $\t=(\eta \, \, \gamma)^T$, we plot the estimated values of $\eta$ and $\gamma$ separately in Figures~\ref{fig:Weibull_scale_RR}-\ref{fig:Weibull_shape_RR_minimax}. We then numerically evaluate the \emph{mean square error} (MSE) of the obtained estimators of scale parameter $\eta$ and shape parameter $\gamma$, and compare them with their corresponding Cramer-Rao lower bound (CRLB)~\CR{[XXX]}. To this end, we numerically evaluate CRLB of scale parameter and shape parameter by Monte-Carlo (MC) simulations. We take $1000$ MC runs for this purpose and the results are shown in the table \ref{table:crlb}

\subsection{Bayes estimator}
To illustrate the Bayes TS approach of Section~\ref{subsec:Bayes}, we study the effect of choosing different priors for the parameter $\t$. In particular, two choices of prior are considered:

\begin{enumerate}
\item Independent uniform priors $\mathcal{U}[1,20]$ for $\eta$ and $\gamma$.

\item Independent \emph{uninformative} priors for $\eta$ and $\gamma$ with density \cite{SUN1997319}
\begin{align} \l{eq:uninf}
f(x) &= \begin{cases}
\displaystyle \frac{c}{x}, & \text{if } a\leq x \leq b \\
0, & \text{otherwise},
\end{cases}
\end{align}
where $c = 1/\ln(b/a)$. Here, we take $a=1$ and $b=20$ for both $\eta$ and $\gamma$.
% \item The density of the prior for $\eta$ and $\gamma$ takes the same form as (2), but we take $a=1, b=2$ for $\alpha$ ,and $a=18, b=20$ for $\gamma$ 
% \item The density of the prior for $\eta$ and $\gamma$ takes the same form as (2), but we take $a=18, b=20$ for $\eta$ ,and $a=1, b=2$ for $\gamma$ 
\end{enumerate}

For each of the above priors, $M_{\t} = 1000$ values of $\t=(\eta, \gamma)^T$ (\ie, $1000$ values of $\eta$ and $\gamma$) are drawn independently. For each such value of $\t$, $N=10000$ observations $y^1, \dots, y^N$ are drawn according to the Weibull distribution parameterized by $\t$.

\begin{figure}[h]
\centering
\includegraphics[width=0.8\linewidth]{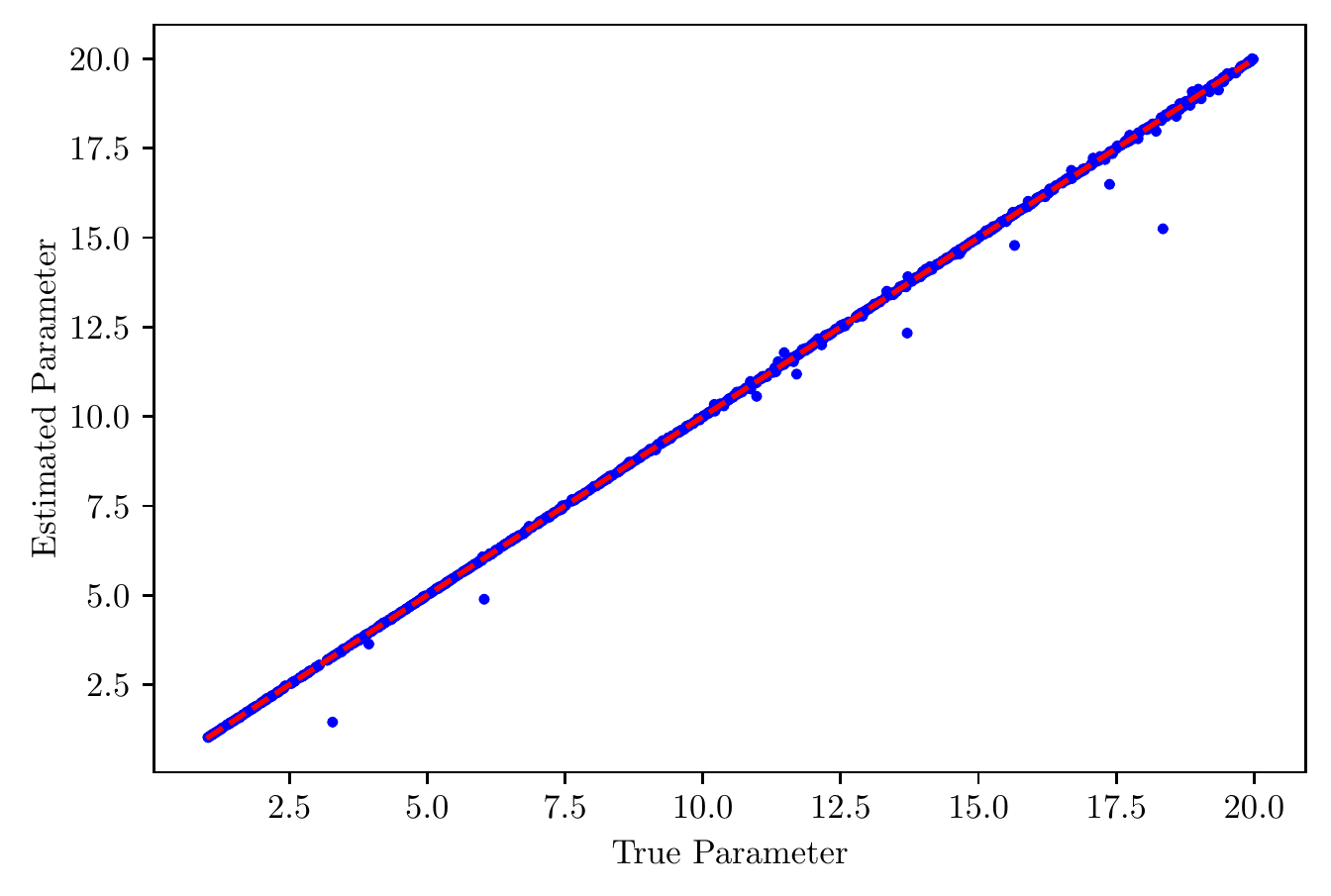}
\caption{Scatter plot (blue dots) of Bayes TS estimates of $\eta$ vs. its true value for a uniform prior. The red dashed line corresponds to an oracle estimate, which knows the true value of the parameter.}
\label{fig:Weibull_scale_RR}
\end{figure}

% \begin{figure}[h]
%     \centering
% \scalebox{0.5}{\input{ts_weibull_iid_scale_param.pgf}}
% \caption{Scatter plot (blue dots) of Bayes TS estimates of $\eta$ vs. its true value for a uniform prior. The red dashed line corresponds to an oracle estimate, which knows the true value of the parameter.}
% \label{fig:Weibull_scale_RR}
% \end{figure}

\begin{figure}[h]
\centering
\includegraphics[width=0.8\linewidth]{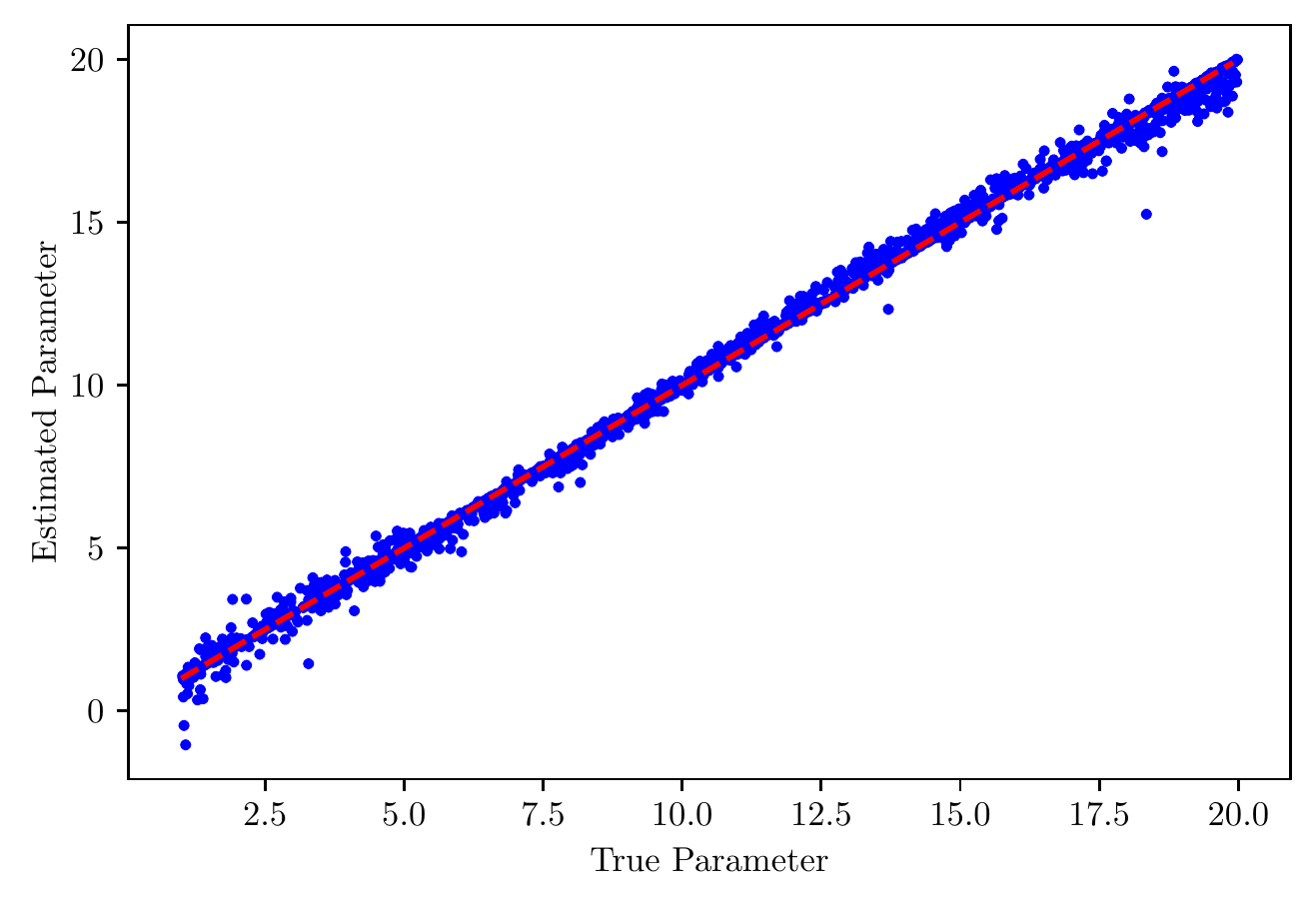}
\caption{Scatter plot (blue dots) of Bayes TS estimates of $\gamma$ vs. its true value for a uniform prior. The red dashed  line corresponds to an oracle  estimate, which knows the true value of the parameter.}
\label{fig:Weibull_shape_RR}
\end{figure}

% \begin{figure}[h]
%     \centering
% \scalebox{0.5}{\input{ts_weibull_iid_shape_param.pgf}}
% \caption{Scatter plot (blue dots) of Bayes TS estimates of $\gamma$ vs. its true value for a uniform prior. The red dashed  line corresponds to an oracle  estimate, which knows the true value of the parameter.}
% \label{fig:Weibull_shape_RR}
% \end{figure}

\begin{figure}[h]
\centering
\includegraphics[width=0.8\linewidth]{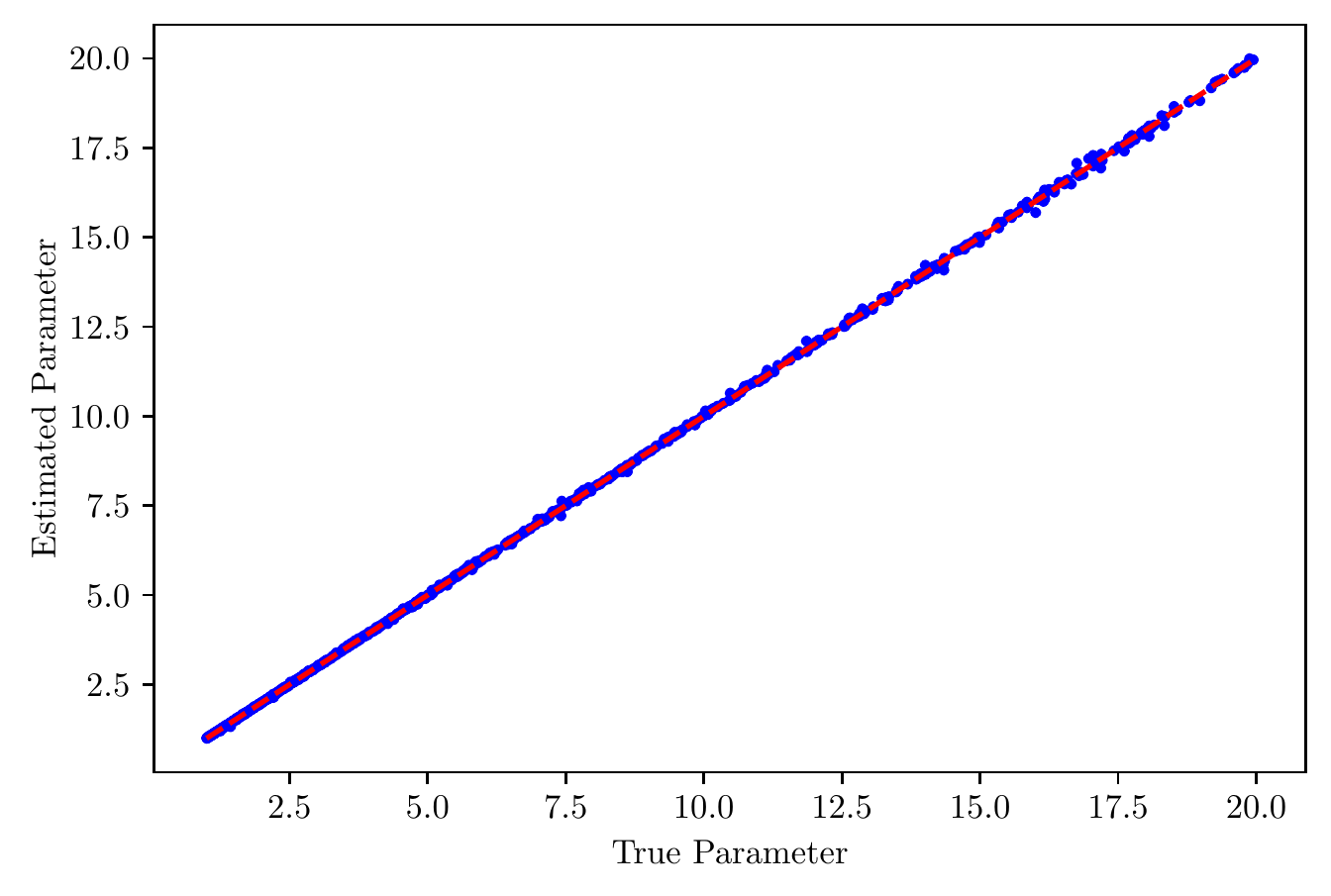}
\caption{Scatter plot (blue dots) of Bayes TS estimates of $\eta$ vs. its true value for an uninformative prior. The red dashed line corresponds to an oracle estimate, which knows  the true value of the parameter.}
\label{fig:Weibull_scale_RR_uninf}
\end{figure}

% \begin{figure}[h]
%     \centering
% \scalebox{0.5}{\input{ts_weibull_iid_scale_param_uninfprior.pgf}}
% \caption{Scatter plot (blue dots) of Bayes TS estimates of $\eta$ vs. its true value for an uninformative prior. The red dashed line corresponds to an oracle estimate, which knows  the true value of the parameter.}
% \label{fig:Weibull_scale_RR_uninf}
% \end{figure}

\begin{figure}[h]
\centering
\includegraphics[width=0.8\linewidth]{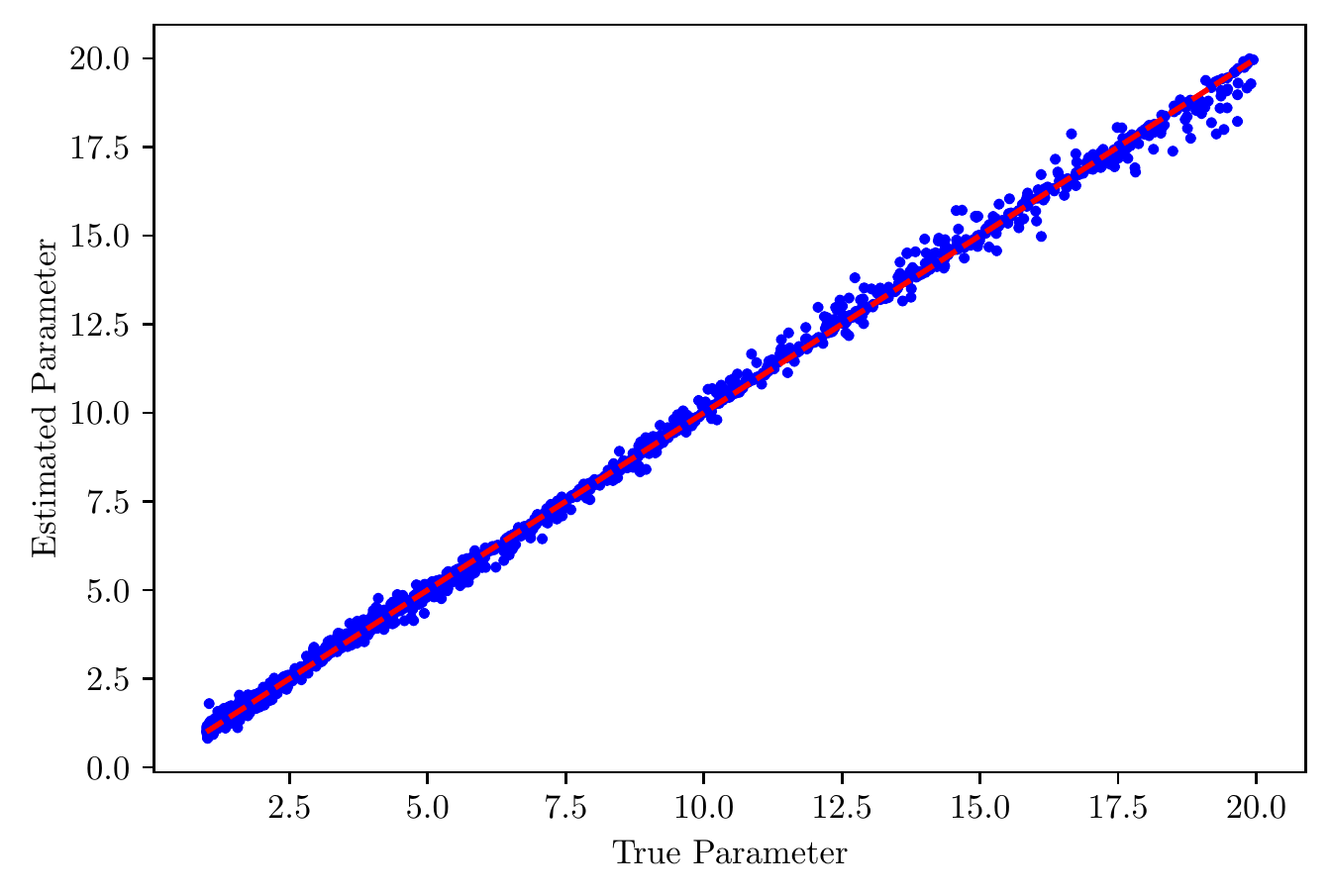}
\caption{Scatter plot (blue dots) of Bayes TS estimates of $\gamma$ vs. its true value for an uninformative prior. The red dashed line  corresponds to an oracle estimate, which knows  the true value of the parameter.}
\label{fig:Weibull_shape_RR_uninf}
\end{figure}

For different choices of prior distribution, we show a scatter plot of estimated values of $\eta$ and $\gamma$ vs. their true values in Figures \ref{fig:Weibull_scale_RR} - \ref{fig:Weibull_shape_RR_uninf}. Figures~\ref{fig:Weibull_scale_RR} and \ref{fig:Weibull_shape_RR} correspond to uniform prior, whereas Figures~\ref{fig:Weibull_scale_RR_uninf} and \ref{fig:Weibull_shape_RR_uninf} correspond to the uninformative prior given by \eqref{eq:uninf}.

\subsection{Minimax estimator}

\begin{figure}[h]
\centering
\includegraphics[width=0.8\linewidth]{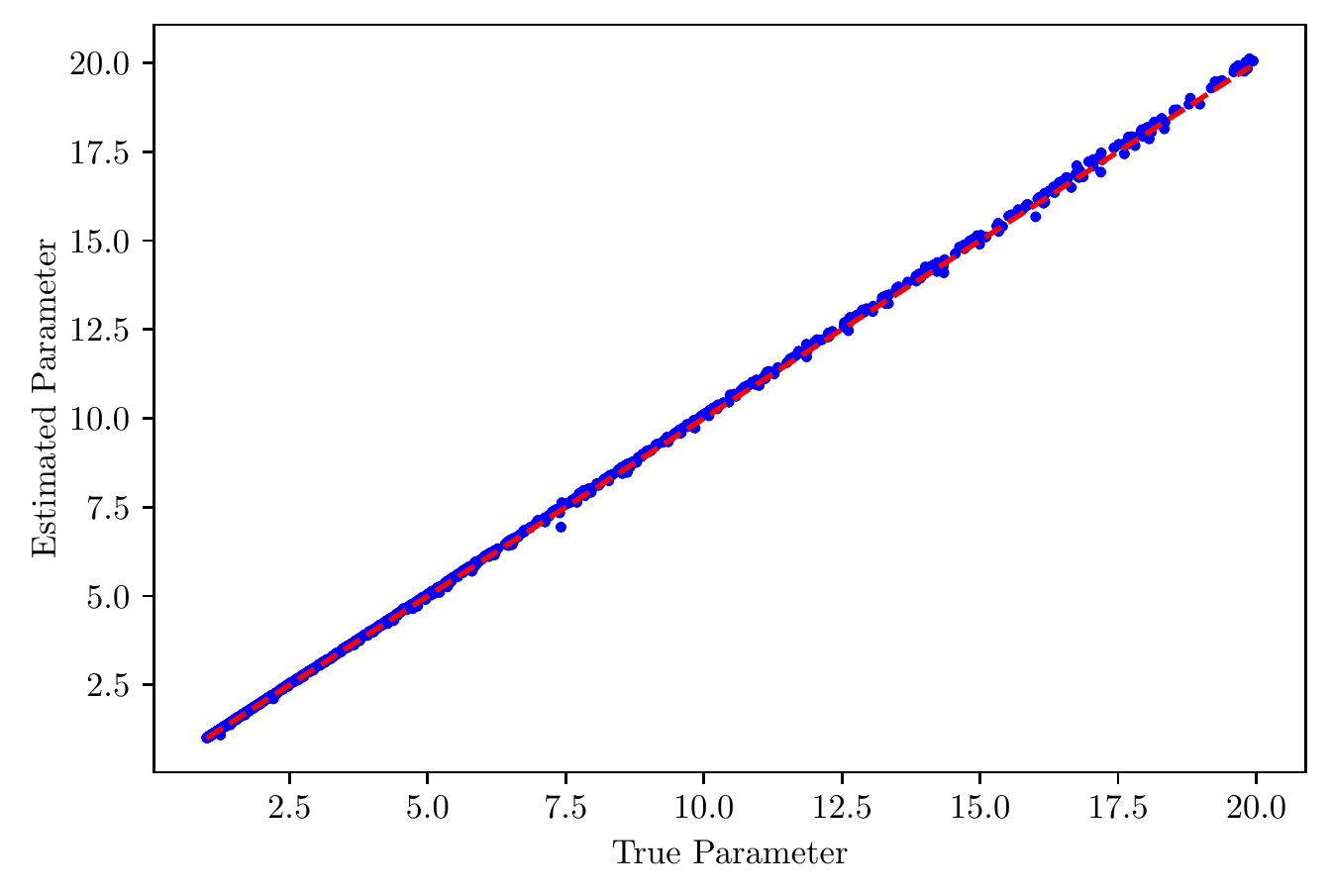}
\caption{Scatter plot (blue dots) of minimax TS estimates of $\eta$ vs. its true value. The red dashed line corresponds to an oracle estimate, which knows the true value of the parameter.}
\label{fig:Weibull_scale_RR_minimax}
\end{figure}

\begin{figure}[h]
\centering
\includegraphics[width=0.8\linewidth]{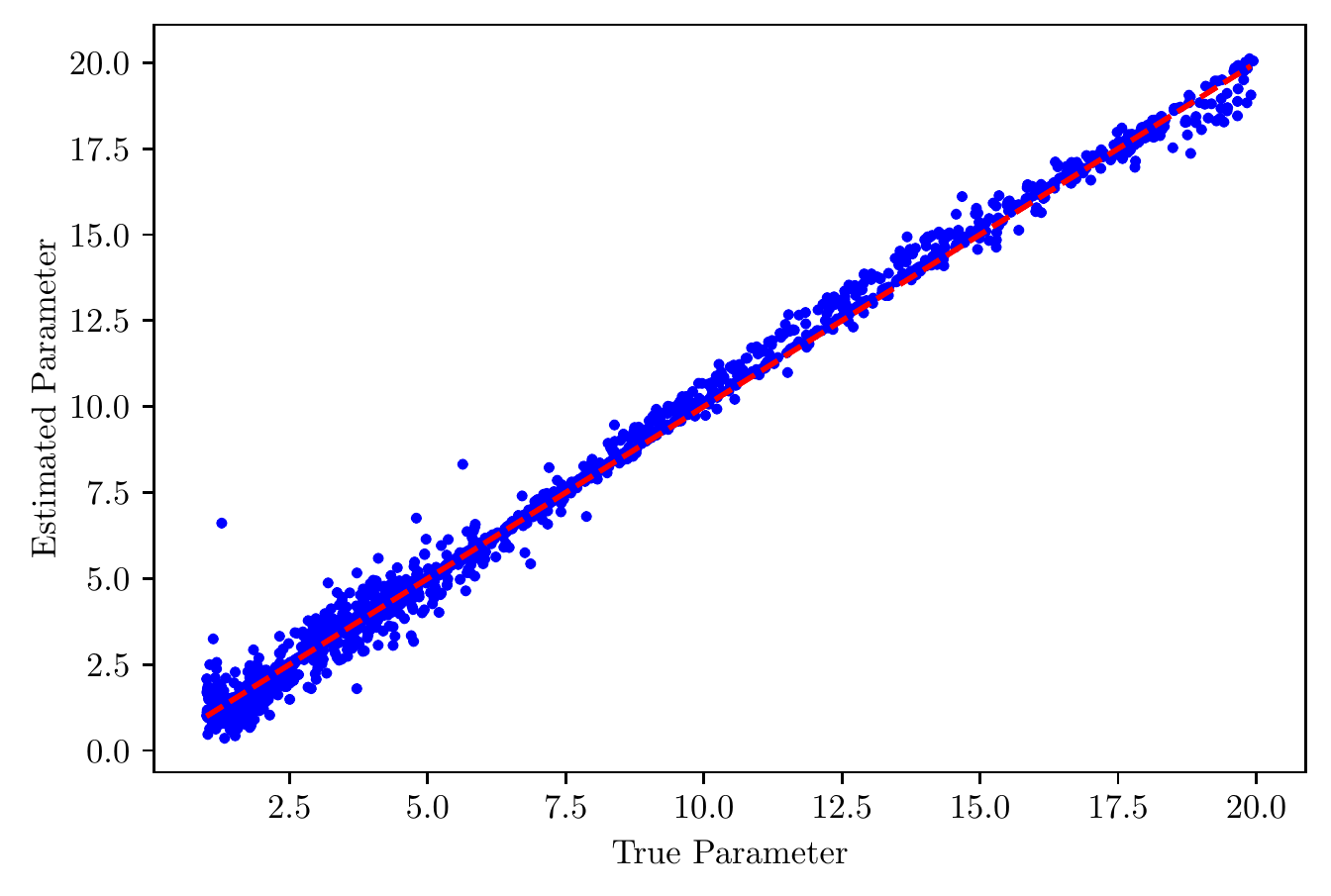}
\caption{Scatter plot (blue dots) of minimax TS estimates of $\gamma$ vs. its true value. The red dashed line corresponds to an oracle estimate, which knows the true value of the parameter.}
\label{fig:Weibull_shape_RR_minimax}
\end{figure}

For the minimax estimator, we take the proposal distribution $s$ to be uniform $\mathcal{U}[1,20]$. To analyse its performance, $M_{\t} = 1000$ values of $\t=(\eta, \gamma)^T$ (\ie, $1000$ values of $\eta$ and $\gamma$) are drawn independently from this proposal distribution. For each such value of $\t$, $N=10000$ observations $y^1, \dots, y^N$ are drawn according to the Weibull distribution parameterized by $\t$.   We show the scatter plot of estimated values of $\eta$ and $\gamma$ vs. their true values in Figures~\ref{fig:Weibull_scale_RR_minimax} and \ref{fig:Weibull_shape_RR_minimax}.

% \begin{figure}[h]
%     \centering
% \scalebox{0.5}{\input{ts_weibull_iid_scale_param_minimax.pgf}}
% \caption{Scatter plot (blue dots) of minimax TS estimates of $\eta$ vs. its true value. The red dashed line corresponds to an oracle estimate, which knows the true value of the parameter.}
% \label{fig:Weibull_scale_RR_minimax}
% \end{figure}

\begin{table*}[t]
\centering
\scalebox{0.8}{
  \begin{tabular}{|c|c|c|c|c|c|c|c|c|c|c|}

    \hline
      \multicolumn{2}{|c|}{\textbf{True Values}} & \multicolumn{2}{|c|}{\textbf{CRLB}} & \multicolumn{2}{c|} {\textbf{MSE, Bayesian (Uniform Prior)}} & \multicolumn{2}{c|} {\textbf{MSE, Bayesian (Uninformative Prior)}} & \multicolumn{2}{c|} {\textbf{MSE, Minimax}} \\
    % \hline
    % \textbf{Inactive Modes} & \textbf{Description}\\
    \cline{1-10}
     \text{$\eta$} & \text{$\gamma$} & \text{${\eta}$} & \text{${\gamma}$} & \text{$\hat{\eta}$} &\text{$\hat{\gamma}$}& \text{$\hat{\eta}$} &\text{$\hat{\gamma}$}& \text{$\hat{\eta}$} &\text{$\hat{\gamma}$} \\
    %\hhline{~--}
    \hline\hline
      $2$ & $2$ & $1.11\times10^{-4}$ & $2.43\times10^{-4}$ & $2.58\times10^{-4}$ & $5.77\times10^{-2}$ & $1.42\times10^{-4}$ & $2.06\times10^{-2}$ & $2.17\times10^{-4}$ & $16\times10^{-2}$ \\ \hline
     %-----------------------------------------------------------------
     $2$ & $8$ & $6.93\times10^{-6}$ & $3.89\times10^{-3}$ & $1.11\times10^{-5}$ &  $5.61\times10^{-2}$ & $1.27\times10^{-5}$ & $4.44\times10^{-2}$ & $4.28\times10^{-4}$ & $13.29\times10^{-2}$ \\ \hline
      $4$ & $2$ & $4.43\times10^{-4}$ & $2.43\times10^{-4}$ & $6.74\times10^{-4}$ &  $1.05\times10^{-1}$ & $6.07\times10^{-4}$ & $2.43\times10^{-2}$ & $8.38\times10^{-4}$ & $14.67\times10^{-2}$\\ \hline
      $4$ & $8$ & $2.77\times10^{-5}$ & $3.89\times10^{-3}$ & $3.84\times10^{-5}$ &  $6.40\times10^{-2}$ & $4.33\times10^{-5}$ & $3.96\times10^{-2}$ & $1.72\times10^{-3}$ & $8.59\times10^{-2}$ \\ \hline
        $8$ & $2$ & $1.77\times10^{-3}$ & $2.43\times10^{-4}$ & $2.26\times10^{-3}$ &  $1.89\times10^{-1}$ & $2.27\times10^{-3}$ & $2.35\times10^{-2}$ & $3.307\times10^{-3}$ & $18.605\times10^{-2}$ \\ \hline
         $8$ & $8$ & $1.11\times10^{-4}$ & $3.89\times10^{-3}$ & $1.58\times10^{-4}$ &  $7.901\times10^{-2}$ & $1.76\times10^{-4}$ & $4.51\times10^{-2}$ & $6.77\times10^{-3}$ & $8.39\times10^{-2}$\\ \hline
  \end{tabular}
}
\caption{MSE of Bayes and minimax TS estimators of the scale and shape parameters, and their corresponding CRLBs.}
\label{table:crlb}
\end{table*}

\bigskip
To assess the quality of the Bayes and minimax TS estimators of $\eta$ and $\gamma$, we have numerically evaluated the \emph{mean square error} (MSE) of the TS estimators, based on $1000$ Monte Carlo simulations, and compared them with a numerical evaluation of their corresponding Cram\'er-Rao lower bounds (CRLB)~\cite{crlbweibull}. The results are shown in Table~\ref{table:crlb}.

\medskip
It is evident from Figures~\ref{fig:Weibull_scale_RR} - \ref{fig:Weibull_shape_RR_minimax} that the estimates given by both Bayes and minimax TS estimators of the scale ($\eta$) and shape ($\gamma$) parameters are close to the true values. However, according to Table~\ref{table:crlb}, the TS estimates of $\eta$ are fairly efficient (in the sense that their MSE is close to CRLB), while those of $\gamma$ are not as reliable. The reason for the latter lies in the choice of the function $\phi$ for $\gamma$.
Also, one can see from Table~\ref{table:crlb} the effect of changing priors in the Bayes estimator: For the \emph{uninformative} priors, estimates of $\eta$ and $\gamma$ are relatively much closer to true values than for the uniform priors. It is also interesting to note that, while no prior is needed for the minimax estimator, one can still get reliable estimates, as demonstrated in Figures~\ref{fig:Weibull_scale_RR_minimax} and \ref{fig:Weibull_shape_RR_minimax}.
By more careful ``feature engineering'' of $\phi$ in the second stage for $\hat{\gamma}$, we could obtain more reliable estimates of the shape parameter without using more complex functional approximators such as Deep Neural Networks. How to perform such feature engineering, based on a theoretical analysis of our novel TS estimators, is left for future research.

\section{Conclusions} \l{sec: conclusions}
In this paper, we provided a decision-theoretical framework for the Two-Stage paradigm to estimate the parameters of a data generating model. In particular, we have derived Bayes and minimax formulations of TS, and showed how to implement them for \emph{i.i.d.} data, by considering nonlinear functions of the sample quantiles of the data as a first stage, and then optimizing for a linear function as the second stage. We have also evaluated the performance of these novel estimators on simulated data from a Weibull distribution. 

% \begin{figure}[h]
%     \centering
% \scalebox{0.6}{\input{ts_weibull_iid_scale_param_minimax.pgf}}
% \end{figure}

% \input{ts_weibull_iid_shape_param_minimax.pgf}

\bibliography{References}
\end{document}